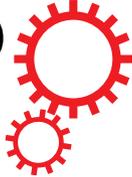

# SCIENTIFIC REPORTS

**OPEN**

# Signal coverage approach to the detection probability of hypothetical extraterrestrial emitters in the Milky Way

Claudio Grimaldi

**The lack of evidence for the existence of extraterrestrial life, even the simplest forms of animal life, makes it is difficult to decide whether the search for extraterrestrial intelligence (SETI) is more a high-risk, high-payoff endeavor than a futile attempt. Here we insist that even if extraterrestrial civilizations do exist and communicate, the likelihood of detecting their signals crucially depends on whether the Earth lies within a region of the galaxy covered by such signals. By considering possible populations of independent emitters in the galaxy, we build a statistical model of the domain covered by hypothetical extraterrestrial signals to derive the detection probability that the Earth is within such a domain. We show that for general distributions of the signal longevity and directionality, the mean number of detectable emitters is less than one even for detection probabilities as large as 50%, regardless of the number of emitters in the galaxy.**

The possibility that extra-solar planets in the Milky Way Galaxy may harbor intelligent, communicating civilizations has long inspired the search for signs of their existence. While data suggest that a significant fraction of Sun-like stars may host Earth-sized planets within the habitable zone[1], the chances of detecting hypothetical extraterrestrial signals still remain highly speculative. Estimates of the detection probability[2–4] often rely on arguments, such as the Drake equation[5], aimed at determining the odds that communicating civilizations exist in the Milky Way. Yet, given the lack of empirical data, it is highly uncertain that even microbiological life exists elsewhere than the Earth[6], let alone complex, intelligent life.

Even if the existence of extraterrestrial broadcasting civilizations is assumed, a necessary condition for their detection is that the Earth falls within a region of space covered by the emitted signals. The fractional volume occupied by the union of such signals is therefore a natural measure of the probability of detection, as in the study of two-phase heterogeneous media[7], where the volume fraction of one phase gives the probability that a randomly chosen point belongs to that phase. In applying this concept to the detection of galactic signals, however, additional factors such as the spatial and age distribution of hypothetical emitters, their distance from the Earth, and the degree of directionality of the emitted signals and their longevity, must be taken into account.

By considering the ensemble of possible realizations of these factors, we build a statistical model of the galactic domain covered by hypothetical extraterrestrial signals and define the detection probability as the probability that the Earth's position intersects such a domain. We shall not deal with the question of the existence of extrasolar life in the galaxy, as we are concerned here exclusively with the communicative phase of hypothetical civilizations. We show that the constraints imposed by the finite size of the galaxy and the speed of light naturally imply that the age scale of the oldest detectable signal, if there is any, is negligible compared to the age scale of possible communicating civilizations in the Milky Way. This observation enables us to derive the detection probability without a full knowledge of the characteristics of the signals. In particular, we show that distributed signal longevities affect the detection probability only through their first moment. By adopting the statistical independence of the signals, we derive a universal upper bound for the mean number ($\bar{k}$) of remotely-detectable extraterrestrial communicating civilizations, which sets a fundamental limit on the number of detectable civilizations in the Milky Way. We show

Laboratory of Physics of Complex Matter, Ecole Polytechnique Fédérale de Lausanne, Station 3, CH-1015 Lausanne, Switzerland. Correspondence and requests for materials should be addressed to C.G. (email: claudio.grimaldi@epfl.ch)

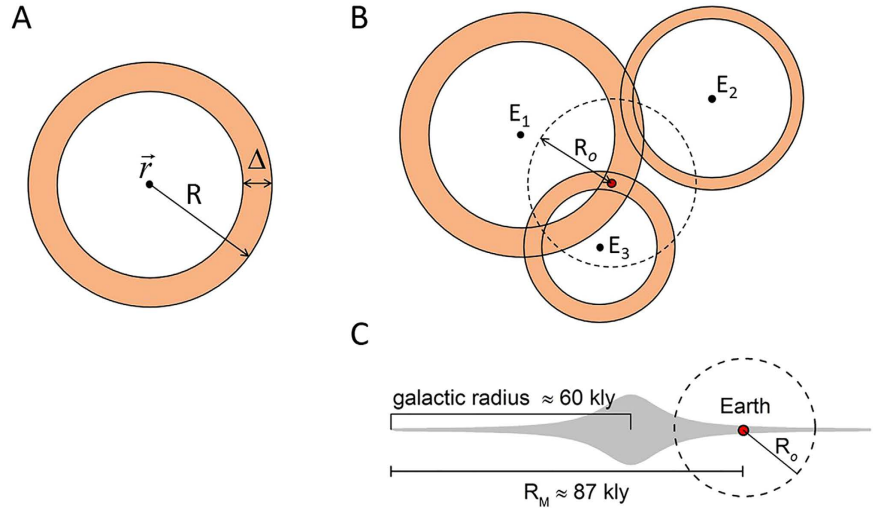

**Figure 1. Detectable domain for isotropic signals.** (**A**) Two-dimensional representation of a spherical shell signal of outer radiur $R$ and shell thickness $\Delta$ emitted from $\vec{r}$. (**B**) The detectable domain formed by union of three spherical shells centered at $E_1$, $E_2$, and $E_3$. In this example the Earth (red dot) intersects the shells of $E_1$ and $E_3$, but only $E_3$ is within the observable radius $R_o$. (**C**) Schematic illustration of the location of the Earth within the Milky Way galaxy. The dashed circle represents the observable sphere of radius $R_o$ centered about the Earth.

that $\bar{k} < 1$ for a detection probability as large as 50%, even if all stars in the Milky Way are scanned for extraterrestrial signals, regardless of the assumed number of hypothetical emitters and of the longevity of their signals.

## Detection probability for isotropic signals

We start by describing in details the formalism used for the special case of isotropic, non-directional electromagnetic signals, as this formalism applies as well to our subsequent generalization to anisotropic, beam-like signal emissions. In the following, we employ the term "signal" to indicate any kind of electromagnetic radiation, regardless of its specific wavelength and power spectrum, emitted from an extraterrestrial civilization located within the Milky Way. The restriction to our galaxy reflects the typical range of current SETI initiatives and, as it is discussed below, it allows for some important simplification in the analysis.

### Single emitter

Let us assume that there exists a communicating civilization (the emitter) located at some point $\vec{r}$ in the Milky Way. The emitter transmits an isotropic signal starting at time $t = -t_0$ and, after a time interval $\Delta t$ during which the signal is active, the emitter ceases to communicate. At the present time ($t = 0$) the region covered by the signal is a spherical shell centered about $\vec{r}$ of outer radius $R = ct_0$ and thickness $\Delta = c\Delta t$, where $c$ is the speed of light[3] (Fig. 1A). The signal is detectable, at least in principle, only if the vector position of the Earth at the present time ($\vec{r}_o$) falls within that spherical shell (Fig. 1B). This condition can be expressed in terms of the indicator function

$$f_{R,\Delta}(\vec{r} - \vec{r}_o) = \begin{cases} 1, & R - \Delta \leq |\vec{r} - \vec{r}_o| \leq R \\ 0, & \text{otherwise}. \end{cases} \quad (1)$$

We make the additional assumption that only signals from emitters within an observable radius $R_o$ from the Earth are searched for. In principle, $R_o$ may represent the distance beyond which the signal becomes indiscernible from the background noise [in that case $R_o$ depends on the luminosity of the signal and the receiver sensitivity[8,9]]. For the moment, however, we shall treat $R_o$ as a free parameter that ranges from a few light-years up to the maximum possible distance $R_M$ of an emitter from the Earth. We estimate $R_M$ as the sum of the galactic radius ($\approx 60$ kly) and the distance of the Earth from the galactic center ($\approx 27$ kly), as shown schematically in Fig. 1C.

Next, we introduce the probability $\rho_E(\vec{r})d\vec{r}$ of finding the emitter within a volume $d\vec{r}$ centered about $\vec{r}$ ($\int d\vec{r} \rho_E(\vec{r}) = 1$, where the integration is extended over the entire galaxy). The odds that the signal is detectable are thus given by the joint probability that the spherical shell signal intersects the Earth and that the emitter is at most at distance $R_o$,

$$p(R_o; R, \Delta) = \int d\vec{r} \rho_E(\vec{r}) \theta(R_o - |\vec{r} - \vec{r}_o|) f_{R,\Delta}(\vec{r} - \vec{r}_o), \quad (2)$$

where $\theta(x) = 1$ if $x \geq 0$ and $\theta(x) = 0$ if $x < 0$. By replacing the indicator function by unity, Eq. (2) gives the probability

$$\pi(R_o) = \int d\vec{r} \rho_E(\vec{r}) \theta(R_o - |\vec{r} - \vec{r}_o|) \quad (3)$$

| Parameters | Signification |
|---|---|
| $R$ | Outer radius of a spherical shell signal |
| $\Delta, \Delta_M, \overline{\Delta}$ | Thickness, maximum thickness, and average thickness of a spherical shell signal |
| $\Omega_0$ | Solid angle of a conical beam signal |
| $R_o$ | Observable radius |
| $RM$ | Maximum possible distance of an emitter from the Earth |
| $t_M = (R_M + \Delta_M)/c$ | Age of the oldest detectable signal |
| $N$ | Number of galactic signals emitted within a time $t_M$ from present |
| $\rho_E(\vec{r})$ | Density probability that an emitter is found at $\vec{r}$ |
| $T^*$ | Time interval spanned by the age distribution of communication civilizations in the Milky Way. |
| $\phi$ | Detection probability (it is defined as the probability that there is at least one emitter within a distance $R_o$ whose signal intercepts the Earth) |
| $\overline{k}$ | Mean number of detectable emitters |

**Table 1. Legend of symbols used in the text and their signification.**

that the emitter is within the observable radius from the Earth. Equation (3) represents therefore the maximum possible $p(R_o; R, \Delta)$ for a given $R_o$.

Since we are interested in the probability of detecting any kind of spherical shell signal, we marginalize Eq. (2) over all possible values of $R$ and $\Delta$. To this end, we make the reasonable assumption that the duration of a signal is uncorrelated with its starting time, implying that $R$ and $\Delta$ are distributed independently according to some probability distribution functions, $\rho_R(R)$ and $\rho_\Delta(\Delta)$. Next, we observe that since the farthest emitter in the Milky Way is at distance $R_M$ ($\approx$87 kly) from the Earth, a spherical shell signal with an inner radius larger than $R_M$ cannot intersect the Earth's orbit, independently of the emitter position in the galaxy. In integrating out $R$ and $\Delta$ in Eq. (2) we thus adopt the condition that $R \leq R_M + \Delta$ to exclude those cases where signals are undetectable in principle. Although it is not strictly necessary (see Supplementary Information), we assume for simplicity that $\rho_\Delta(\Delta)$ is a bounded distribution with the maximum allowed shell thickness $\Delta_M$. We therefore define the detection probability of an isotropic signal as

$$p(R_o) = \frac{\int_0^{\Delta_M} d\Delta \rho_\Delta(\Delta) \int_0^{R_M+\Delta} dR \rho_R(R) p(R_o; R, \Delta)}{\int_0^{\Delta_M} d\Delta \rho_\Delta(\Delta) \int_0^{R_M+\Delta} dR \rho_R(R)}, \tag{4}$$

which represents the conditional probability that the Earth intersects a spherical shell signal from an emitter within a radius $R_o$, given that the signal has been emitted within a time $t_M = (R_M + \Delta_M)/c$ before present (see Table 1 for symbol legend).

Actually, we do not need detailed knowledge of $\rho_R(R)$ and $\rho_\Delta(\Delta)$ to calculate Eq. (4). To understand this point, we note that $\rho_R(R)$ varies on a length-scale of order $cT^*$, where $T^*$ is the time scale possibly spanned by the age distribution of *communicating* civilizations in the Milky Way[10]. We speculate that $T^*$ is of the order of a few Gy, which results from the difference between the epoch at which the first habitable planets in the Milky Way formed [$\approx$8–9 Gy before present[11]] and the time elapsed from the formation of the Earth to the emission of the first radio and TV signals ($\approx$4.5 Gy), which we take as a proxy. Monte Carlo simulations of the evolution of intelligent life in the galaxy[12,13] support similar estimates of $T^*$. Therefore, even for $\Delta_M$ extending up to several million light-years, we expect that the age of the oldest detectable signal is such that $t_M \ll T^*$, which implies that $\rho_R(R)$ does not show appreciable variations over a length-scale of order $R_M + \Delta_M = ct_M$. This key observation allows us to approximate $\rho_R(R)$ by a constant, which cancels out of Eq. (4), and to write the detection probability as

$$p(R_o) = \frac{\overline{\Delta}}{R_M + \overline{\Delta}} \pi(R_o), \tag{5}$$

where $\overline{\Delta} = \int_0^{\Delta_M} d\Delta \Delta \rho_\Delta(\Delta)$ is the mean thickness of the spherical shell (equivalently, $\overline{\Delta}/c$ is the mean signal longevity). The above expression is remarkable because it shows that the detection probability is simply proportional to $\pi(R_o)$, with a factor of proportionality that depends on $\rho_\Delta(\Delta)$ only through its first moment. From Eq. (5) we see also that a signal does not need to last billions of years to be detectable with a significant probability[14], as a mean signal longevity larger than about $R_M/c \approx 87$ ky suffices to make $p(R_o)$ essentially equivalent to $\pi(R_o)$.

To assess the effect of the emitter density probability on $p(R_o)$, we take as a proxy for $\rho_E(\vec{r})$ the distribution of stars in the galactic habitable zone (GHZ) of the Milky Way[11,15–17]. For simplicity, we consider GHZ models with cylindrical symmetry, and take $\rho_E(\vec{r})$ to have the following expression in cylindrical coordinates [$\vec{r} = (r, \varphi, z)$]:

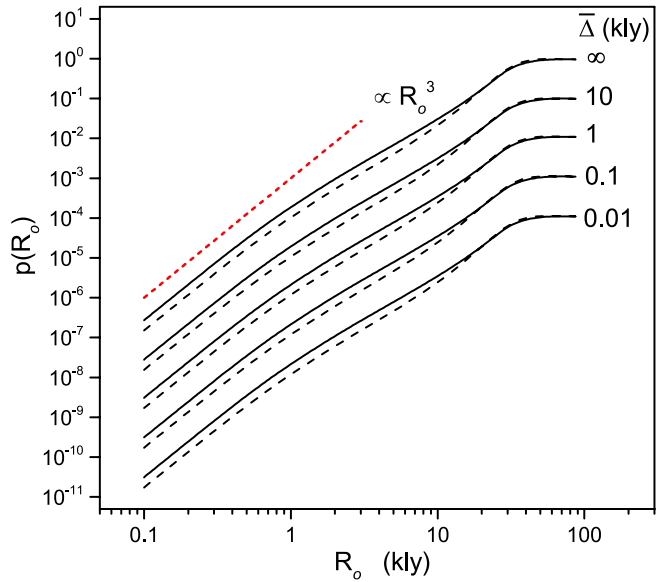

**Figure 2. The detection probability $p(R_o)$ is plotted as a function of the observable radius $R_o$.** The calculations are done by using the emitter distribution function $\rho_E(\vec{r})$ given in Eq. (6). Solid lines are the results for $\rho_E(\vec{r})$ being proportional to the distribution of stars in the galactic thin disk, while the dashed lines are the results for an emitter distribution concentrated on an annular region of the disk comprised between ≈15 kly and ≈34 kly from the galactic center. Each pair of solid and dashed lines represents $p(R_o)$ for the values of the mean shell thickness $\overline{\Delta}$ indicated in the figure. For $\overline{\Delta} = \infty$, $p(R_o)$ reduces to the probability $\pi(R_o)$ (Eq. (3)) of finding an emitter within a radius $R_o$ from the Earth.

$$\rho_E(\vec{r}) = \rho_0 r^m e^{-r/r_s - |z|/z_s}, \qquad (6)$$

where $m$ is an integer, $r$ is the radial distance from the galactic center, $z$ is the height from the galactic plane, and $\rho_0 = [4\pi z_s r_s^{m+2}(m+1)!]^{-1}$ is a normalization constant. We choose the parameters $m$, $r_s$, and $z_s$ in Eq. (6) so as to simulate two representative models of the GHZ. In the first model we assume that the GHZ extends over the entire galactic thin disk[16,17]. Following[18], we take $m = 0$, $r_s = 8.15$ kly, and $z_s = 0.52$ kly. In the second model we reproduces the main features of the annular GHZ of ref. 11 by taking $m = 7$, $r_s = 3.26$ kly, and $z_s = 0.52$ kly. These parameters give a probability of 68% of finding stars with the highest potential to harbor complex life within ≈15 kly and ≈34 kly from the galactic center, which is a good match with the time averaged GHZ of ref. 11. Figure 2 shows that the extent of the observable radius strongly affects the detection probability, with little dependence on the form chosen for $\rho_E(\vec{r})$. In particular, $p(R_o) \propto R_o^3$ for values of $R_o$ smaller than the thickness of the galactic thin disk (≈2 kly) because in the galactic neighborhood the emitter probability distribution is fairly homogeneous. For $R_o \gtrsim 20$ kly the particular form of $\rho_E(\vec{r})$ has practically no effect on the detection probability. These features persist also if we increase the annulus of the emitter distribution from the center to the periphery of the galaxy, although a stronger variation is observed at fixed $R_o$ (see Supplementary Information).

### Multiple emitters

To generalize the detection probability to the case of multiple emitters, we assume that they are independently and identically distributed in space with probability density $\rho_E(\vec{r})$ and that their signals cover spherical shells with independent and identically distributed outer radii and thicknesses. Given $N$ such signals in the entire galaxy, each emitted since a time $t_M$ before present and having detection probability $p(R_o)$, the probability $P(k)$ that the Earth intersects $k$ signals from emitters within a radius $R_o$ is given by a binomial distribution:

$$P(k) = \binom{N}{k} p(R_o)^k [1 - p(R_o)]^{N-k}. \qquad (7)$$

We define the detection probability as $\phi = 1 - P(0)$, which is the probability that there is at least one emitter within the observable radius whose signal intercepts the Earth. Therefore, from Eqs (5) and (7) we find:

$$\phi = 1 - \left[1 - \frac{\overline{\Delta}}{R_M + \overline{\Delta}} \pi(R_o)\right]^N. \qquad (8)$$

Equation (8) has far-reaching consequences regarding the population of hypothetical emitters in the galaxy and the likelihood of detecting their communications. This is best illustrated in Fig. 3, where we show the number of emitters obtained from Eq. (8) for given values of $\phi$. As the observable radius $R_o$ diminishes, the emitter population in the galaxy increases to keep the detection probability constant. The enhancement of the total emitter population as $R_o$ decreases is such that, for $\phi$ larger than about 1%, $N$ can be comparable or even larger than the number of planets ($\approx 10^8$) recently claimed to possibly harbor complex life[19]. Note, however, that $N$, by construction, represents the galactic population of emitters that have sent signals only within a time $t_M$ before present. Even assuming that $\approx 10^8$ is a plausible estimate for the number of planets sustaining complex life and that a significant fraction (10%) of those planets have developed communicating civilizations in the last Gy ($\approx T^*$), $N$ cannot exceed approximately $0.1 \times 10^8 t_M/T^* \approx 10^3$ for $t_M \approx 100$ ky. In this case we infer from Fig. 3 that signals from the galactic neighborhood ($R_o \lesssim 1$ kly) have detection probabilities smaller than 0.01% for a mean signal longevity of 100 years ($\overline{\Delta} = 0.1$ kly). Significantly larger values of $\phi$ can be attained by enhancing the observable radius so as to comprehend the entire galaxy or assuming values of $\overline{\Delta}$ so large that they become comparable to $R_M$.

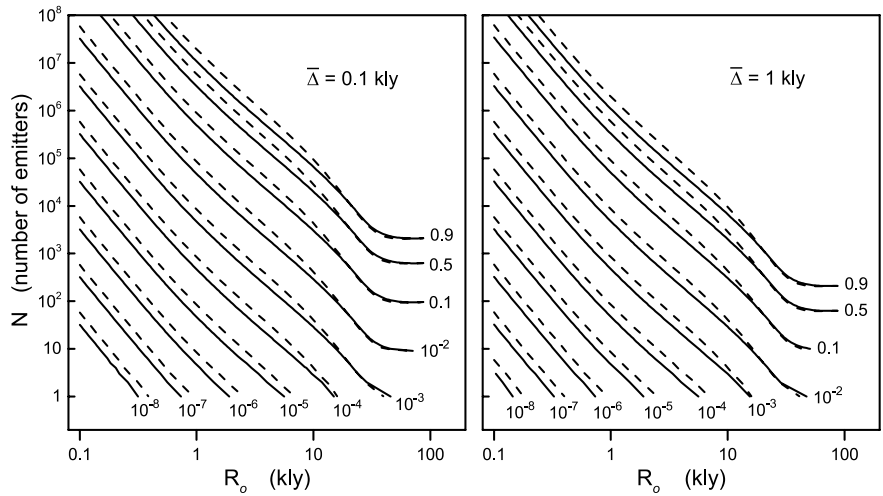

**Figure 3. Population of emitters in the Milky Way.** Calculated number $N$ of emitters in the galaxy as a function of the observable radius for different detection probabilities $\phi$. Each pair of solid and dashed lines represents $N$ for the values of $\phi$ indicated in the figure. The solid and dashed lines have the same meaning as in Fig. 2.

It is worth to stress at this point that the value of $N$ extracted from the likelihood $\phi$ that the Earth is within the galactic domain covered by the signals (as done above) is not directly comparable to the number (say, $N_D$) of communicative civilizations appearing in the Drake equation[5]. In its original formulation, indeed, the Drake equation estimates $N_D$ from a product of probabilities that events, which are necessary for the development of communicative life in the galaxy, occur regardless of whether or not the emitted signals can be detected in principle. Later reformulations of the Drake equation follow a similar route[20–22]. We note also that the signal longevity $\overline{\Delta}/c$ enters explicitly both in the present formulation and in the Drake equation (where it is commonly denoted by the symbol $L$). However, as it was remarked in the previous section, its effect on the detection probability saturates at values larger than about $R_M/c \approx 87$ ky because the size of the galaxy and the speed of light imply an upper bound to the age of the oldest detectable signal.

### Mean number of detectable emitters

Equation (8) gives the likelihood that the Earth intersects a region of detectable signals, independently of whether or not we look for such signals. Signal detection would occur only if we point our detectors at emitters whose signals intercept the Earth at the time of measurement. The mean number ($\overline{k}$) of such signals is therefore the quantity of direct interest in the search for extraterrestrial signals and it is easily calculated from the degree probability distribution of Eq. (7), which gives: $\overline{k} = \sum_{k=0}^{N} k P(k) = N \overline{\Delta}/(R_M + \overline{\Delta}) \pi(R_o)$. By using the detection probability of Eq. (8), we can conveniently rewrite $\overline{k}$ as

$$\overline{k} = N[1 - (1-\phi)^{1/N}], \qquad (9)$$

where the distribution of the emitters, the observable radius, and the mean signal longevity are encoded in $\phi$, which can be regarded as an input (albeit unknown) parameter. The advantage of expressing $\overline{k}$ in terms of $\phi$ is highlighted in Fig. 4, which shows $\overline{k}$ as a function of the detection probability for different values of $N$: $\overline{k}$ depends weakly on $N$ for any chosen values of $\phi$ smaller than about 0.9–0.95. More importantly, for any $N$ the mean number of detectable signals is bounded from above as

$$\overline{k} \leq \ln\left(\frac{1}{1-\phi}\right), \qquad (10)$$

where the equality corresponds to the Poisson distribution of an infinite number of signals. Equation (10) sets therefore a fundamental limit on $\overline{k}$ that is independent of the number of hypothetical extraterrestrial civilizations in the galaxy and of the characteristics of the isotropic signals. As discussed in more details in the following sections, Eq. (10) is a robust result because it relies only on the assumption of statistical independence of the signals, without invoking arguments about the existence and number of extrasolar communicating civilizations. One striking implication of Eq. (10) is that for detection probabilities as large as about 50–60%, for which the number of potentially detectable emitters in the galaxy can reach values well beyond $\approx 10^3$ even for signals lasting in average 1′000 years (see Fig. 3), the mean number of detectable signals is below one. The region covered by hypothetical extraterrestrial signals can thus comprehend a significant fraction of the galaxy without us even noticing it. This is perhaps the most compelling argument that the so-called Fermi paradox is, actually, not a paradox.

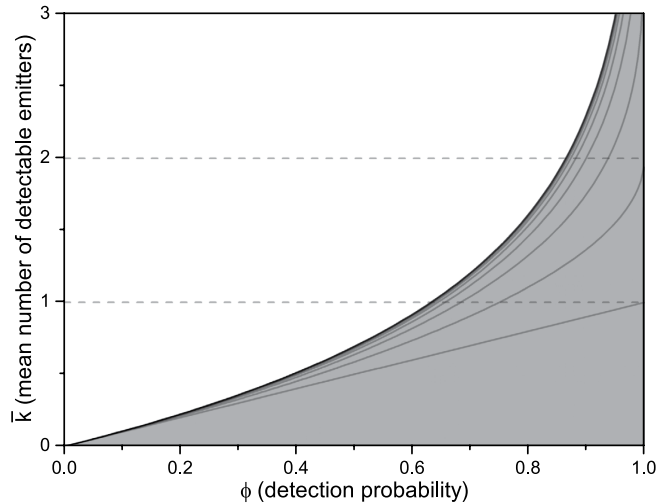

**Figure 4. Solid lines represent the mean number $\overline{k}$ of detectable signals as a function of the detection probability $\phi$ for different numbers $N$ of emitters in the galaxy.** $N = 1, 2, 4, 8, 16, \ldots$ from bottom to top. The thick solid line is $\ln(1-\phi)^{-1}$ while the grey region in the figure indicates $\overline{k} < \ln(1-\phi)^{-1}$.

### Detection probability of anisotropic (beam-like) signals

We can now extend our analysis to the case in which the emitters send anisotropic, directional signals, such as collimated beams[23,24]. We discuss here two limiting situations[2]: one in which all beams are directed towards the Earth, in which case the signals are intentional, and the other in which the beams are directed in random directions in space. In the latter case the signals are unintentional and the detection would be accidental.

Let us start by consider an emitter at position $\vec{r}$ that transmits, starting at a time $t_0$ before present and during a time interval $\Delta t$, a conical beam signal of solid angle $\Omega_0$ and axis oriented along the direction of a unit vector $\hat{n}$. At present time, the region covered by the beam is given by the intersection between a spherical shell centered at $\vec{r}$ of outer radius $R = ct_0$ and shell thickness $\Delta = c\Delta t$, and a cone with axis directed along $\hat{n}$, vertex centered at $\vec{r}$, and solid angle $\Omega_0 = 2\pi(1 - \cos\gamma_0)$, where $\gamma_0$ is the apex semi-angle (Fig. 5).

The condition that the Earth intersects this region is given in terms of the indicator function:

$$f_{R,\Delta}(\vec{r} - \vec{r}_o; \hat{n}) = \theta\left(\hat{n} \cdot \frac{\vec{r}_o - \vec{r}}{|\vec{r}_o - \vec{r}|} - \cos\gamma_0\right) f_{R,\Delta}(\vec{r} - \vec{r}_o), \tag{11}$$

where $\vec{r}_o$ is the vector position of the Earth and $f_{R,\Delta}(\vec{r} - \vec{r}_o)$ is the indicator function of the spherical shell as given in Eq. (1). For a given $\hat{n}$, the probability that the beam intersects the Earth and the emitter is at most at distance $R_o$ is given by

$$p_b(R_o; R, \Delta, \hat{n}) = \int d\vec{r} \rho_E(\vec{r}) \theta(R_o - |\vec{r} - \vec{r}_o|) f_{R,\Delta}(\vec{r} - \vec{r}_o, \hat{n}). \tag{12}$$

In the case of an intentional signal the beam is directed towards the Earth. The $\theta$-function in Eq. (11) becomes 1 because $\hat{n} \cdot (\vec{r}_o - \vec{r})/|\vec{r}_o - \vec{r}| = 1$, and Eq. (12) reduces to the probability that the Earth intersects a spherical shell signal of outer radius $R$ and shell thickness $\Delta$, as in Eq. (2). It follows therefore that the results obtained for the case of isotropic (spherical-shell) signals are equally valid for beams directed towards the Earth. In particular, the detection probability of $N$ independent beams directed towards the Earth coincides with that of $N$ isotropic spherical shell signals considered above (Eq. (8)) where now $\overline{\Delta}$ represents the beam mean length.

In the case in which the beam axis is oriented at random, the probability of intersection (Eq. (12)) must be mediated over all orientations of $\hat{n}$. From Eqs (11) and (12) we find:

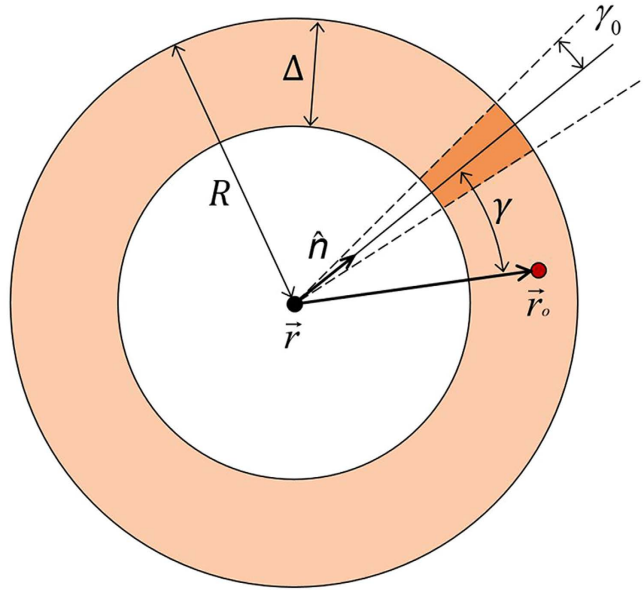

**Figure 5. Two dimensional scheme of the volume covered by a conical beam signal with axis directed along the unit vector $\hat{n}$.**

$$\begin{aligned} p_b(R_o; R, \Delta, \Omega_0) &= \int \frac{d\hat{n}}{4\pi} p_b(R_o; R, \Delta, \hat{n}) \\ &= p(R_o; R, \Delta) \int_0^\pi \frac{d\gamma}{2} \sin\gamma\, \theta(\cos\gamma - \cos\gamma_0) \\ &= \frac{\Omega_0}{4\pi} p(R_o; R, \Delta), \end{aligned} \quad (13)$$

where $\gamma$ is the angle between $\hat{n}$ and $\vec{r}_o - \vec{r}$, and $p(R_o; R, \Delta)$ is the probability that the Earth intersects a spherical shell of outer radius $R$ and shell thickness $\Delta$ (Eq. (2)). From Eq. (13) we see that an highly collimated random beam can dramatically reduce the detection probability. Let us consider for example the case in which $\Omega_0$ is chosen so that at distance $R_b$ from the emitter the beam covers a given area $A$. This area can be chosen to represent, for example, the size of a planetary system targeted by the beam and located at distance $R_b$ from the emitter. In this case, Eq. (13) represents the likelihood that the Earth intersects a collimated signal that was not intended for us. From $\Omega_0 = A/R_b^2$, and denoting $d$ the diameter of the targeted planetary system, we obtain $\Omega_0/4\pi \approx (d/4R_b)^2$ for $d < R_b$. Taking $d$ comparable to the size of the solar system ($\approx 60$ AU $\approx 10^{-3}$ ly, if we delimit the solar system by the orbit of Neptune) we see that $\Omega_0/4\pi$ can be as small as $\approx 6 \times 10^{-8}$, $\approx 6 \times 10^{-10}$, and $\approx 6 \times 10^{-12}$ for $R_b = 1$ ly, 10 ly, and 100 ly, respectively. Note that the assumption of random beam orientations is justified as long as $R_b$ is much smaller than the thickness of the galactic thin disk.

By following the same steps as described earlier for the case of multiple isotropic signals, we obtain that the detection probability $\phi$ of $N$ independent and randomly oriented beams is:

$$\phi = 1 - \left[1 - \frac{\overline{\Omega}_0}{4\pi} \frac{\overline{\Delta}}{R_M + \overline{\Delta}} \pi(R_o)\right]^N, \quad (14)$$

where $\overline{\Delta}$ is the mean beam length and, if we assume that the beam solid angles are independently and identically distributed, $\overline{\Omega}_0$ is the mean solid angle. Using Eq. (14), it is straighforward to show that the mean number $\overline{k}$ of detectable beams satisfies Eq. (9) and the inequality in Eq. (10).

### Detection probability of a general combination of independent signals

We conclude our analysis by showing that the inequality (10) holds true also for a general combination of independent isotropic and anisotropic signals. To make the derivation as general as possible, we shall drop the assumption that the duration and the starting time of a signal are uncorrelated and identically distributed by introducing a generalized distribution function $\rho_i(R, \Delta, \Omega_0)$ for the $i$-th signal, where $\Omega_0$ denotes the solid angle. We shall also consider the possibility that the observable radius depends in general on $\Delta$ and $\Omega_0$ as these quantities may be related to the strength and the power spectrum of the signal. For example, decreasing the solid angle of a beam-like signal increases the strength of the signal and consequently it increases the observable radius. Also, a longer signal longevity may be ascribed to a more advanced civilization which transmits stronger signals. The probability of intersection of a generic signal can be written therefore as

$$p(R, \Delta, \Omega_0) = \frac{\Omega_0}{4\pi} p[R_o(\Delta, \Omega_0); R, \Delta], \qquad (15)$$

where $\Omega_0 = 4\pi$ for isotropic or beam-like signals directed towards the Earth and $\Omega_0 \ll 4\pi$ for beams oriented at random. By adopting the condition that $R \leq R_M + \Delta$, the conditional probability of Eq. (4) generalizes to:

$$p_i = \frac{\int_0^\infty d\Delta \int_0^{R_M+\Delta} dR \int_0^{4\pi} d\Omega_0 \rho_i(R, \Delta, \Omega_0) p(R, \Delta, \Omega_0)}{\int_0^\infty d\Delta \int_0^{R_M+\Delta} dR \int_0^{4\pi} d\Omega_0 \rho_i(R, \Delta, \Omega_0)}, \qquad (16)$$

where the maximum shell thickness $\Delta_{M,i}$ of the $i$-th signal is encoded in $\rho_i(R, \Delta, \Omega_0)$. Note that even for the general case of Eq. (16) we can safely neglect the dependence on $R$ of $\rho_i(R, \Delta, \Omega_0)$ because $(R_M + \Delta_{M,i})/c$ is expected to be negligible with respect to the age scale of communicating civilizations in the galaxy. For a collection of $N$ signals emitted within a time $t_M = (R_M + \Delta_M)/c$ before present, where now $\Delta_M$ denotes the maximum among all $\Delta_{M,i}$, the probability $P(k_1, ..., k_N)$ that the Earth intersects $k_i = 0, 1$ signals of type $i$ ($i = 1, 2, ..., N$) is

$$P(k_1, ..., k_N) = \prod_{i=1}^N p_i^{k_i} (1 - p_i)^{1-k_i}, \qquad (17)$$

from which we obtain the probability that the Earth intersects at least one signal

$$\phi = 1 - P(0, ..., 0) = 1 - \prod_{i=1}^N (1 - p_i), \qquad (18)$$

and the mean number of detectable signals

$$\overline{k} = \sum_{k_1=0,1} \cdots \sum_{k_N=0,1} (k_1 + \cdots + k_N) P(k_1, ..., k_N) = \sum_{i=1}^N p_i. \qquad (19)$$

To derive a relation between $\phi$ and $\overline{k}$, we note that Eq. (18) satisfies $\ln(1 - \phi) = \sum_{i=1}^N \ln(1 - p_i) < \sum_{i=1}^N p_i$, which, using Eq. (19), gives directly the inequality of Eq. (10), which sets therefore a universal upper bound for the mean number of detectable signals, valid for any combination of independent isotropic and anisotropic signals from the Milky Way.

## Conclusions

Using the fractional volume of the galaxy occupied by hypothetical extraterrestrial electromagnetic signals, we derive the probability that the Earth intercepts at least one of such signals from communicating civilizations in the galaxy. In contrast to previous studies, the present approach focuses primarily on the coverage properties of the signals rather than the possible number of galactic emitters. In this way, we are able to determine the detection probability $\phi$ for isotropic (non-directional) and anisotropic (directional) signals, and for any combination of them. For a given $\phi$, we derive a universal upper bound for the mean number of detectable signals $\overline{k}$ (Eq. (10)) which is independent of the emitter population in the galaxy. The present results may have profound implications for the search for extraterrestrial communicating civilizations and and they may affect our perception of the likelihood of detecting them, as implied by the finding that $\overline{k}$ is below unity even if we assume that a significant fraction of the galaxy is covered by extraterrestrial signals. An interesting corollary to this result is that in the hypothetical event that an extraterrestrial signal is actually detected, the number of galactic emitters that have broadcasted a signal within the last $t_M \approx 100'000$ years is likely to be large.

Our results can be improved by allowing the observable radius to depend on the signal spectra and by better modeling the galactic habitable zone[17,25]. For example, interstellar medium may sensibly reduce the extent of the observable radius $R_o$ in the radio wavelength range, but leave $R_o$ for optical emissions relatively unaffected. Recent simulations show that the GHZ, which we use as a proxy for the emitter probability distribution, does not display the simple cylindrical symmetry used here[25] and that globular clusters may present favorable conditions for the development of intelligent life[26]. Although these factors may alter the functional dependence of the detection probability, the inequality (10) remains unaffected as long as the statistical independence of the signals is assumed.

## References

1. Petigura, E., Howard, A. & Marcy, G. Prevalence of earth-size planets orbiting sun-like stars. *Proc. Natl. Acad. Sci.* **110,** 19273–19278 (2013).
2. Horvat, M. Calculating the probability of detecting radio signals from alien civilizations. *Int. J. Astrobiol.* **5,** 143–149 (2006).
3. Smith, R. Broadcasting but not receiving: density dependence considerations for seti signals. *Int. J. Astrobiol.* **8,** 101–105 (2009).
4. Wandel, A. On the abundance of extraterrestrial life after the kepler mission. *Int. J. Astrobiol.* **14,** 511–516 (2015).
5. Drake, F. The radio search for intelligent extraterrestrial life. In *Current Aspects of Exobiology* 323–345 (Pergamon Press, New York, 1965).
6. Spiegel, D. & Turner, E. Bayesian analysis of the astrobiological implications of life's early emergence on earth. *Proc. Natl. Acad. Sci.* **109,** 395–400 (2012).
7. Torquato, S. *Random Heterogeneous Materials: Microstructure and Macroscopic Properties* (Springer, New York, 2002).
8. Townes, C. At what wavelengths should we search for signals from extraterrestrial intelligence? *Proc. Natl. Acad. Sci.* **80,** 1147–1151 (1983).
9. Wilson, T. The search for extraterrestrial intelligence. *Nature* **409,** 1110–1114 (2001).

10. Hair, T. Temporal dispersion of the emergence of intelligence: an inter-arrival time analysis. *Int. J. Astrobiol.* **10,** 131–135 (2011).
11. Lineweaver, C., Fenner, Y. & Gibson, B. The galactic habitable zone and the age distribution of complex life in the milky way. *Science* **303,** 59–62 (2004).
12. Forgan, D. & Nichol, R. A failure of serendipity: the square kilometre array will struggle to eavesdrop on human-like extraterrestrial intelligence. *Int. J. Astrobiol.* **10,** 77–81 (2011).
13. Morrison, I. & Gowanlock, M. Extending galactic habitable zone modeling to include the emergence of intelligent life. *Astrobiology* **15,** 683–696 (2015).
14. Grinspoon, D. *Lonely Planets: The Natural Philosophy of Alien Life* (HarperCollins, New York, 2003).
15. Gonzalez, G., Brownlee, D. & Ward, P. The galactic habitable zone: Galactic chemical evolution. *Icarus* **152,** 185–200 (2001).
16. Prantzos, N. On the "galactic habitable zone". *Space Sci. Rev.* **135,** 313–322 (2008).
17. Gowanlock, M., Patton, D. & McConnell, S. A model of habitability within the milky way galaxy. *Astrobiology* **11,** 855–873 (2011).
18. Misiriotis, A., Xilouris, E., Papamastorakis, J., Boumis, P. & Goudis, C. The distribution of the ism in the milky way-a three-dimensional large-scale model. *Astron. Astrophys.* **459,** 113–123 (2006).
19. Irwin, L., Méndez, A., Fairén, A. & Schulze-Makuch, D. Assessing the possibility of biological complexity on other worlds, with an estimate of the occurrence of complex life in the milky way galaxy. *Challenges* **5,** 159–174 (2014).
20. Ćirković, M. M. The Temporal Aspect of the Drake Equation and SETI. *Astrobiology* **4,** 225–231 (2004).
21. Maccone, C. The statistical Drake equation. *Acta Astronaut.* **67,** 1366–1383 (2010).
22. Glade, N., Ballet, P. & Bastien, O. A stochastic process approach of the Drake equation parameters. *Int. J. Astrobiol.* **11,** 103–108 (2016).
23. Schwartz, R. & Townes, C. Interstellar and interplanetary communication by optical masers. *Nature* **190,** 205–208 (1961).
24. Forgan, D. Can collimated extraterrestrial signals be intercepted? *J. Br. Interplanet. Soc.* **67,** 232–236 (2014).
25. Forgan, D., Dayal, P., Cockell, C. & Libeskind, N. Evaluating galactic habitability using high-resolution cosmological simulations of galaxy formation. *Int. J. Astrobiol.* **16,** 60–73 (2017).
26. Di Stefano, R. & Ray, A. Globular clusters as cradles of life and advanced civilizations. *Astrophys. J.* **827,** 54 (2016).

### Acknowledgements

The author thanks Avik P. Chatterjee for helpful comments.

### Author Contributions

C.G. conceived the study, analyzed the results and wrote the manuscript.

### Additional Information

**Supplementary information** accompanies this paper at http://www.nature.com/srep

**Competing Interests:** The author declares no competing financial interests.

**How to cite this article:** Grimaldi, C. Signal coverage approach to the detection probability of hypothetical extraterrestrial emitters in the Milky Way. *Sci. Rep.* **7,** 46273; doi: 10.1038/srep46273 (2017).

**Publisher's note:** Springer Nature remains neutral with regard to jurisdictional claims in published maps and institutional affiliations.